\documentclass[twocolumn,superscriptaddress,floatfix,preprintnumbers]{revtex4}
\usepackage{graphics,amssymb,amsmath,epsfig,color}
\usepackage{epstopdf}
\usepackage{graphicx}

\begin{document}

\title{Magneto-conductivities in Weyl semimetals: the effect of chemical potential and temperature}
\author{Xiao Xiao$^1$, K. T. Law}
\affiliation{Department of Physics, The Hong Kong University of Science and Technology, Hong Kong, People¡¯s Republic of China}
\author{P. A. Lee}
\email{palee@mit.edu}
\affiliation{Department of Physics, Massachusetts Institute of Technology, Cambridge, Massachusetts 02139, USA}
\date{\today}
\begin{abstract}
  We present the detailed analyses of magneto-conductivities in a Weyl semimetal within Born and self-consistent Born approximations. In the presence of the charged impurities, the linear magnetoresistance can happen when the charge carriers are mainly from the zeroth (n=0) Landau level. Interestingly, the linear magnetoresistance is very robust against the change of temperature, as long as the charge carriers mainly come from the zeroth Landau level. We denote this parameter regime as the high-field regime. On the other hand, the linear magnetoresistance disappears once the charge carriers from the higher Landau levels can provide notable contributions. Our analysis indicates that the deviation from the linear magnetoresistance is mainly due to the deviation of the longitudinal conductivity from the $1/B$ behavior. We found two important features of the self-energy approximation: 1. a dramatic jump of $\sigma_{xx}$, when the $n=1$ Landau level begins to contribute charge carriers, which is the beginning point of the middle-field regime, when decreasing the external magnetic field from high field; 2. In the low-field regime $\sigma_{xx}$ shows a $B^{-5/3}$ behavior and results the magnetoresistance $\rho_{xx}$ to show a $B^{1/3}$ behavior. The detailed and careful numerical calculation indicates that the self-energy approximation (including both the Born and the self-consistent Born approximations) does not explain the recent experimental observation of linear magnetoresistance in Weyl semimetals.
\end{abstract}
\maketitle

\section{introduction}

Properties of gapless Dirac fermions attract much attention in the community, since the successful fabrication of monolayer graphene \cite{Novoselov04}. Its dispersion spectrum is characterized by two Dirac points at which the conduction band and valance band touch. Importantly, The Dirac points in graphene are stabilized by symmetries with a $\pm \pi$ Berry phase, as long as both the time reversal and inversion symmetry are present. Due to the Dirac dispersion spectrum, graphene has been a novel platform to explore interesting physics in two-dimension, such as the novel properties of collective modes \cite{Neto09}, unconventional effects of interactions and disorders \cite{Neto09,Sarma11,Kotov12}, and the exciting properties under magnetic field \cite{Zhang05,Goerbig11}.

Very recently, three-dimension Dirac Fermions are predicted \cite{Wang12, Wang13} to be harbored in Na$_{3}$Bi and Cd$_{3}$As$_{2}$, and it was verified later in the angle-resolved photoemission spectroscopy (ARPES) experiments \cite{Xu15a}. The Dirac semimetal phase in these materials is stabilized by the time reversal and the inversion symmetry together with crystalline symmetry \cite{Yang14}, so the low energy physics around the Dirac point is described at least by a $4\times4$ matrix and the conduction and valence bands are two-fold degeneracy. When the time reversal or inversion symmetry is broken, the two Dirac points in these materials would be split into two pairs of Weyl points with opposite chirality. In this sense, the Weyl semimetal is robust and does not depends on any symmetry (except for the translation symmetry in the crystal). Recently, ARPES experiments had identified the Weyl semimetal phase in TaAs \cite{Xu15b} and NbAs \cite{Xu15c}. In addition to materials belong to the same family of TaAs and NbAs \cite{Weng15}, the pyrochlore iridates \cite{Wan11} and topological insulator heterostructures \cite{Burkov11} are predicted to harbor Weyl femrions. To conserve the total chirality, the Weyl points should appear in pairs with opposite chirality, which thus leads to the chiral anomaly (or the Adler-Bell-Jackiw anomaly) \cite{Adler69, Bell69}. It indicates that the electrons can be pumped from one node to the other with opposite chirality at a rate proportional to $\vec{E}\cdot\vec{B}$ \cite{Aji12,Son13,Liu13,Parameswaran14}, when applying the electric field parallel to the external magnetic fields. Therefore, the observation of the negative magnetoresistance is expected to be a signature of chiral anomaly \cite{Huang15}. However, there are unexpected phenomena under external magnetic field in addition to the chiral anomaly. Very strikingly and interestingly, in the recent experiments \cite{Liang15,Feng15,Novak15,Zhao15}, in the configuration that the magnetic field is perpendicular to the electric field, a linear and non-saturated magnetoresistence is observed in the Dirac semimetals. This is in contrast to the standard relaxation time analysis of the Boltzmann equation, where the resistivity saturates for $\omega_{c}\tau\gg1$. In the above $\omega_{c}$ is the cyclotron frequency and $\tau$ is the relaxation time.

The early work of Abrikosov was among the few examples where linear magnetoresistance was predicted in a single band system \cite{Abrikosov98}. Abrikosov showed that in a Dirac electronic spectrum, when the chemical potential coincides with the zeroth Landau level, linear magnetoresistance is obtained in the Born approximation in a model with long range changed impurities scattering. Recent papers \cite{Klier15,Pesin15} have re-examined this problem in details with the self-consistent Born approximation. In particular, ref.\cite{Klier15} showed that Coulomb impurity is crucial for the linear magnetoresistance found by Abrikosov. Meanwhile, Song \textit{et al} \cite{Song15} approached the problem from the point of view of the classical motion of guiding center in the case of random potential which is slowly varying in space. In the case where the cyclotron orbit size is smaller than the distance scale of the variation of the random potential, they showed that linear magnetoresistance is obtained even when the chemical potential is far away from the Dirac neutrality point.

Motivated by these experimental observations and the theoretical progresses \cite{Lu17}, in this work we calculate the magneto-conductivities in a Weyl semimetal in the presence of charged impurities, which are crucial for the appearance of linear magnetoresistance within the self-energy approximation adopted here \cite{Abrikosov98,Klier15,Pesin15}. Different from the previous work \cite{Klier15,Pesin15}, where the authors focus on the discussion of the case with $\mu=0$ and the role played by the zeroth Landau level, we focus on the situation that the chemical potential is away from the $\mu=0$ neutrality point to answer the question whether the linear magnetoresistance can happen in this regime within the self-energy approximations. We have to notice that in experiments the chemical potential in Weyl semimetal samples is usually away from $0$, and thus not only the zeroth Landau level but also the higher Landau levels with index $n\geq1$ should play important roles. Consequently, to understand the roles played by the higher Landau levels with $n\geq1$ is central for our study. For this purpose, we assume that the system is with a given carrier density, and then the chemical potential can be tuned by the external magnetic field. When the magnetic field is large enough, the chemical potential locates in the zeroth Landau level. In this parameter regime, the charge carriers are mainly from the zeroth Landau level, and the system enters the so-called quantum limit regime. Our calculations shows that both Hall and longitudinal conductivities are inversely proportional to magnetic field ($1/B$ behaviors), and thus the system shows the linear magnetoresistance. This result is consistent with the previous studies by using both Born and self-consistent Born approximations\cite{Abrikosov98,Klier15}. Moreover, we show that the $1/B$ behaviors of the conductivities are very robust against the change of temperature, as long as the chemical potential is kept in this regime. On the other hand, by decreasing the magnetic field, the chemical potential can cross higher Landau levels. When this happens, the Hall conductivity shows a slight deviation from the $1/B$ behavior. In the contrast, the change of longitudinal conductivity is dramatic and temperature dependent. When the temperature is low (much smaller than the typical gap between Landau levels $T \ll \sqrt{v^2eB/c}$), the change of the longitudinal conductivity with the decreasing magnetic field shows a `step-like' structure. Moreover, due to the high density of states at the edges of Landau levels, a local maximum in longitudinal conductivity appears accordingly. When the temperature is comparable to $\sqrt{v^2eB/c}$, the longitudinal conductivity shows a universal monotonous power law increasing with the decreasing magnetic field ($\sim B^{-5/3}$). Consequently, the magnetoresistance shows a $B^{1/3}$ behaviors in this parameter regime. However, for the experimental situations the chemical potential is generally far from the neutrality point, and the magnetoresistance $\rho_{xx}$ increases with the magnetic field with the scaling $\sim B$ monotonously (small deviations from $\sim B$ may happen in measurements). We therefore conclude that the self-energy approximation can not explain the experimental observations of linear magnetoresistance in Weyl semimetals. Presumably the results of Song \textit{et al} \cite{Song15} are in a limit where the Born approximation is not valid, thus explaining the different conditions of the two different approaches.

Our discussions would be divided into three sections. In the following sectoin (section II), we will describe a general model of a Weyl semi-metal. Based on the analysis of the model Hamiltonian, we introduce the Born and self-consistent Born approximations to handle the charged impurities. The key point in the calculation of the self-energy is to correctly evaluate the overlapping integrals between Landau levels. This is only done approximately in the existing literatures \cite{Abrikosov98,Klier15,Pesin15}. Their approximations are only valid, when few Landau levels are involved. Therefore, to achieve our purpose, we propose a general scheme to evaluate the overlapping integral to the desired accuracy. The formulation of conductivities and chemical potential are outlined in the last parts of the section. In the section III, we first present the numerical results for the behaviors of $\mu$ and the Hall conductivity $\sigma_{xy}$, which are independent of the self-energy approximation. Then we discuss the longitudinal conductivity $\sigma_{xx}$. In the last part of this section, we make a comparison between the calculated magnetoresistance and an available experimental data to show where the self-energy approximation fails. In the last section of the manuscript (section IV), a summary of our findings is concluded.

\section{Model and approximations}

Our discussion begins with the low energy effective Hamiltonian for a single Weyl node. In this section, we explain the theoretical model of the Weyl semi-metal and how the self-energy approximations are applied to study the magnetoresistance. First we give explanations for the Hamiltonian of the system. In the literatures, for this model two different representations are usually used, one of them is adopted by Abrikosov \cite{Abrikosov98}, and the other is adopted recently by Klier \textit{et al} and others \cite{Klier15,Pesin15}. We explicitly demonstrate that the two different representations are identical, and thus the choice of the representation is just for convenience. Further we describe the machinery of self-energy approximations (including both Born and self-consistent Born approximations) to treat the charged impurities. Finally, we briefly describe the formulas for the calculations of the magneto-conductivities.

\subsection{Hamiltonian and Matsubara Green's functions}

Usually the Weyl nodes should appear in pairs, but it is still valid to consider only one Weyl node at a time, if the energy scale of interest is much lower than the Lifshitz point. Then under the external magnetic field the system is described by the following effective Hamiltonian:
\begin{equation}
H\left(\vec{k}\right) = \int d^{3}r \psi^{\dag} \left(\vec{r}\right) v \vec{\sigma} \cdot \left(\vec{k} - \frac{e}{c}\vec{A}\right) \psi \left(\vec{r}\right),
\end{equation}
where $\vec{\sigma}$ is the Pauli matrices, $\vec{A}$ is the vector potential induced by the external magnetic field, and $v$ is the Fermi velocity. We assume that the external magnetic field is along the z-direction, and choose the gauge that the vector potential takes the form $\vec{A}=\left(0,Bx,0\right)$.

There are two different representations for the Green's function of the model. The first one is to construct the Green's function from the eigenstates of the Hamiltonian Eq.(1) \cite{Abrikosov98}. Actually, from the Hamiltonian Eq.(1) the Landau level energies can be solved:
\begin{equation}
E_{n}^{\pm} = \pm v \sqrt{k_{z}^{2}+2n/\ell_B^2},
\end{equation}
where we have defined the magnetic length as $\ell_B=(eB/c)^{-1/2}$. The corresponding Landau level wave functions are given by:
\begin{align}
& \psi_{n}^{+}=\sqrt{\frac{1}{2}}\left(\begin{array}{c} \chi_{n}^{+}\left(k_{z},\ell_B^{-2}\right) \phi_{n} \\-i \chi_{n}^{-}\left(k_{z},\ell_B^{-2}\right) \phi_{n-1}  \end{array}\right), \nonumber \\
& \psi_{n}^{-}=\sqrt{\frac{1}{2}}\left(\begin{array}{c} \chi_{n}^{-}\left(k_{z},\ell_B^{-2}\right)\phi_{n} \\ i\chi_{n}^{+}\left(k_{z},\ell_B^{-2}\right)\phi_{n-1} \end{array}\right),
\end{align}
where
\begin{equation}
\begin{cases}
\chi_{n}^{+}\left(k_{z},\ell_B^{-2}\right)=\sqrt{\left(1+\frac{k_{z}}{\sqrt{k_{z}^{2}+2n\ell_B^{-2}}}\right)}, \\
\chi_{n}^{-}\left(k_{z},\ell_B^{-2}\right)=\sqrt{\left(1-\frac{k_{z}}{\sqrt{k_{z}^{2}+2n\ell_B^{-2}}}\right)},
\end{cases}
\end{equation}
and $\phi_{n}\left(\tilde{x}\right)$ is the usual eigenfunction of harmonic oscillator and of the form:
\begin{equation}
\phi_{n}\left(\tilde{x}\right) = \sqrt{\frac{\ell_B^{-1}}{\left(2^{n}n!\right)\sqrt{\pi}}}\exp\left(\frac{-\ell_B^{-2}}{2\tilde{x}^{2}}\right) H_{n}\left(\ell_B^{-1} \tilde{x}\right).
\end{equation}
In the above, we have defined $\tilde{x}=x-k_{y}\ell_B^2$. Then we can define the Matsubara Green's function as \cite{Abrikosov98}:
\begin{align}
&\tilde{G}^{(E)}\left(k_{z},k_{y},x,x',\omega_{m}\right) \nonumber \\
= &\sum_{n,\gamma=\pm}\frac{\psi_{n}^{\gamma}\left(x-k_{y}\ell_B^2\right) \psi_{n}^{\gamma~\dag}\left(x'-k_{y}\ell_B^2\right)}{i\omega_{m}+\mu-E_{n}^{\gamma}\left(k_{z}\right)},
\end{align}
where $\tilde{G}^{(E)}$ denotes taht the Green's function matrix is under the representation of the eigenstates of the original Hamiltonian.

On the other hand, The Green's function can be expressed under the basis of the wave-function of Landau levels, for example:
\begin{equation}
|\Phi_n\rangle = \frac{1}{\sqrt{2}} \left(\begin{array}{c} |\phi_n\rangle \\ |\phi_{n-1}\rangle \end{array}\right),
\end{equation}
where $\langle \tilde{x} | \phi_n\rangle = \phi_n(\tilde{x})$ is defined in Eq.(5). Then Green's function under this representation is just as:
\begin{equation}
\hat{G}^{(L)}(i\omega_n) = \sum_{n,m} |\Phi_n\rangle \langle \Phi_n| \frac{1}{i\omega_n - \hat{H}} |\Phi_m\rangle \langle \Phi_m|,
\end{equation}
where $\hat{G}^{(L)}$ is to denote that the Green's function under the representation of Laudau levels. Actually, $\langle \Phi_n| \frac{1}{i\omega_n - \hat{H}} |\Phi_m\rangle$ in the equation above can be evaluated explicitly as:
\begin{align}
&\langle \Phi_n| \frac{1}{i\omega_n - \hat{H}} |\Phi_m\rangle \nonumber \\
=  &\left( \begin{array}{cc} (i\omega_n-vk_z) \delta_{n,m} & -\frac{i\sqrt{2n} v}{\ell_B} \delta_{n,m-1} \\  \frac{i\sqrt{2n} v}{\ell_B} \delta_{n-1,m} & (i\omega_n+vk_z) \delta_{n-1,m-1} \end{array} \right)^{-1}.
\end{align}
At the same time, we notice that $\langle \Psi_n| \frac{1}{i\omega_n - \hat{H}} |\Psi_m\rangle$ can be diagonalized by:
\begin{align}
\begin{cases}
|\Psi_{n,+}\rangle &= \frac{1}{\sqrt{2}} \left(\begin{array}{cc} \chi_n^+(k_z,\ell_B) & 0 \\ 0 & -i\chi_n^-(k_z,\ell_B) \end{array}\right) |\Phi_n\rangle \nonumber \\
&= \mathcal{M}_{n,+} |\Phi_n\rangle, \\
|\Psi_{n,-}\rangle &= \frac{1}{\sqrt{2}} \left(\begin{array}{cc} \chi_n^-(k_z,\ell_B) & 0 \\ 0 & -i\chi_n^+(k_z,\ell_B) \end{array}\right) |\Phi_n\rangle \nonumber \\
&= \mathcal{M}_{n,-} |\Phi_n\rangle,
\end{cases}
\end{align}
and $\langle \tilde{x}|\Psi_{n,\pm}\rangle = \psi_{n}^{\pm}$, which are defined in Eq.(3). Thus, the two representations are related by unitary transformations defined by matrices $\mathcal{M}_{n,\pm}$. Using the transformations, it can be shown explicitly:
\begin{align}
\hat{G}^{(L)}(i\omega_m) &= \sum_{n} \langle x,k_y |\Phi_n\rangle \langle \Phi_n |i\omega_n-H|\Phi_n\rangle \langle \Phi_n | x,k_y \rangle \nonumber \\
&= \frac{\psi_{n}^+(\tilde{x})\psi_{n}^{+~\dag}(\tilde{x})}{i\omega_m+\mu-E_{n}^{+}} + \frac{\psi_{n}^-(\tilde{x})\psi_{n}^{-~\dag}(\tilde{x})}{i\omega_m+\mu-E_{n}^{-}}.
\end{align}
This is explicitly identical to Eq.(6), and thus the two representations are identical. The identification between the two representations gives us the freedom to choose either of them to do the calculation for the convenience.

\begin{figure}[!th]
  \centering
  \includegraphics[width=0.8\columnwidth]{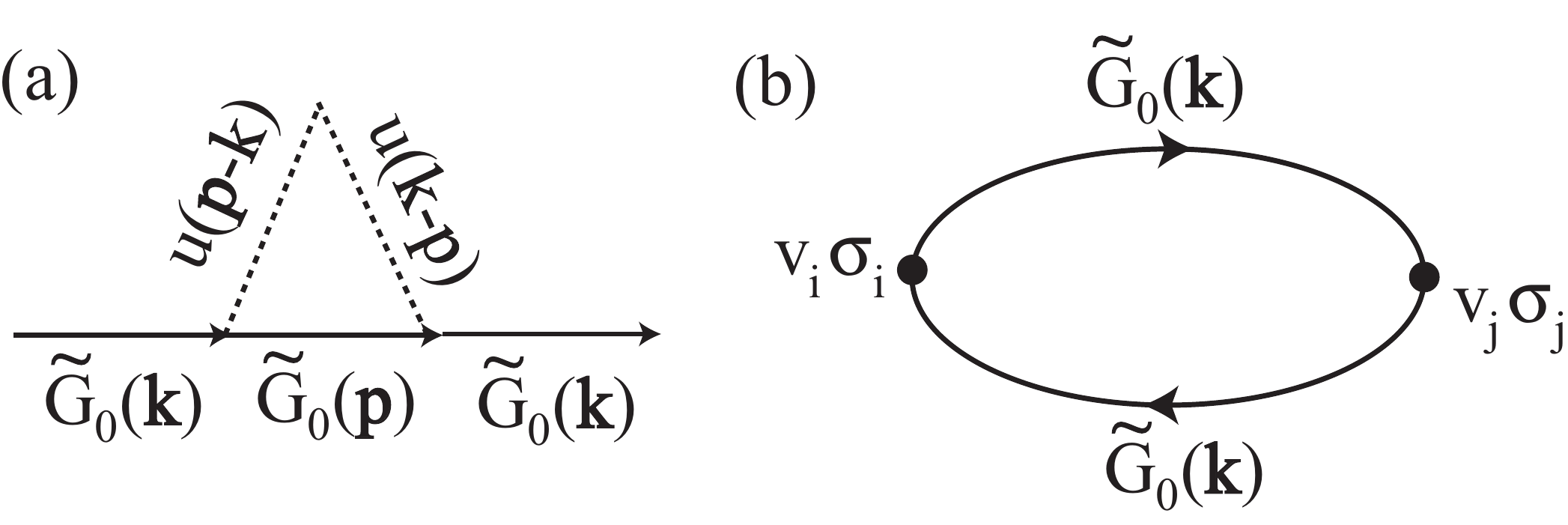}
  \caption{(a) the diagrammatic expression of self-energy approximation due to the charge impurities; (b) the diagrammatic expression of the magneto-conductivity $\sigma_{ij}$. In the figure, the solid line denotes the Green's function, the dashed line denotes the impurity scattering, and the dot denotes the velocity vertex.}
   \label{fig:urtraj1}
\end{figure}

\subsection{Self-energy approximations}

Here we are only interested in the presence of the charged impurities, because in the previous studies the linear magnetoresistance happens only with this kind of impurities. In this case, the impurity scattering potential is given by:
\begin{equation}
u\left(\vec{k}\right) = \frac{4\pi e^{2}}{\varepsilon_{\infty}\left(k^{2}+\kappa^{2}\right)},
\end{equation}
where $\varepsilon_{\infty}$ indicates the background dielectric constant, and the inverse of the screening length $\kappa$ is given by:
\begin{align}
\kappa^{2} &= \frac{4\pi e^{2}}{\varepsilon_{\infty}} \frac{1}{V} \sum_{n,\lambda} \int_{0}^{\beta} d\tau \left\langle T_{\tau} \varrho_{n}^{\lambda}\left(\tau\right)(\varrho^{\lambda}_{n})^{\dag}\left(0\right)\right\rangle \nonumber \\
& = \frac{2e^2\beta}{\pi \varepsilon_{\infty}} \sum_{n,\lambda} \int dk_{z} \frac{dn_{n}^{\lambda}\left(vk_{z}\right)}{d\mu},
\end{align}
where $\varrho_{n}^{\lambda}$ is the density operator for the $\lambda n$ Landau level, and $n_{n}^{\lambda}\left(vk_{z}\right)$ is the Fermi-Dirac distribution for the $\lambda n$ Landau level. The effect of the disorder would enter the self-energy of the Green's functions. We have to note that at the neutrality point ($\mu=0$) the screening length is possibly affected by the external magnetic field \cite{Pesin15}. However, in the present work we focus on the situation $\mu$ away from the neutrality point, and also the strength of disorder is assumed to be small (much smaller than the thermal energy). Therefore, we expect that the modification of the screening length is not important.

To account this effect, we adopted so-called Born approximation, which can be represented by the diagram shown in Fig.1(a). It can be generally written as:
\begin{equation}
\Sigma_n(k_z,i\omega_n) = \sum_{m} \int\frac{dq_z}{2\pi}\mathcal{I}_{n,m}(q_z) \tilde{G}_m^{(0)}(i\omega_n,k_z+q_z),
\end{equation}
in real space. In the above, $\mathcal{I}_{n,m}(q_z)$ describes the overlapping integral between different Landau levels or different eigenstates, and $\tilde{G}_m^{(0)}$ is the single-particle Green's function under either representation mentioned above.

This Born approximation can be improved by further requiring the self-consistency. This self-consistent Born approximation is done by substituting the single-particle Green's function $\tilde{G}_m^{(0)}$ in Eq.(13) by the full Green's function:
\begin{equation}
\tilde{G}_m^{(F)}(i\omega_n) = \left((\tilde{G}_m^{(0)}(i\omega_n))^{-1} + \Sigma_m(i\omega_n) \right)^{-1}.
\end{equation}

Since the two representations are related by a unitary transformations, we just need to work out the case under the Landau level representation. Under this representation, the overlapping integral is written as:
\begin{widetext}
\begin{equation}
\mathcal{I}_{n,m}(q_z) = \frac{1}{8\pi^2\ell_B^2} \int_{|q|\leq1} d\tilde{q}_x d\tilde{q}_y |U(q)|^2 \left| \mathcal{J}_{n,m}(\tilde{q}_x,\tilde{q}_y) + \mathcal{J}_{n-1,m-1}(\tilde{q}_x,\tilde{q}_y) \right|^2,
\end{equation}
\end{widetext}
where:
\begin{equation}
\mathcal{J}_{n,m}(\tilde{q}_x,\tilde{q}_y)=\int d\tilde{x} \phi_n(\tilde{x}) e^{i\tilde{q}_x \tilde{x}} \phi_m(\tilde{x}-\tilde{q}_y).
\end{equation}
In Eq.(15) and (16), we already normalize the in-plane momentum with respect to the magnetic length. In the literature, people are interested in the situation where only few Landau levels play roles, and Eq.(15) can be evaluated approximately. For example, Abrikosov \cite{Abrikosov98} assumed that only the neighbored Landau level have non-vanishing $\mathcal{I}_{n,m}$, while Klier \cite{Klier15} assumed that the overlapping integral is a quantity independent of Landau level index.

For our purpose, we have to evaluate the overlapping integrals as accuracy as possible so that we can still obtain the self-energy correctly, when there are many Landau levels playing roles. To solve this problem, based on the structure of Eq.(15), we can expand the integral with respect to the (normalized) in-plane momentum $\tilde{q}_x$ and $\tilde{q}_y$. The expansion is converging when the expansion order is high enough. This procedure finally arrive at a very convenient expression:
\begin{widetext}
\begin{equation}
\mathcal{I}_{n,m}(\tilde{q}_z) = \frac{2e^4 \ell_B^2}{\epsilon_{\infty}^2} \sum_{j,l,g,h} \left[\frac{1}{2(\tilde{q}_z^2+\tilde{\kappa}^2)(1+(\tilde{q}_z^2+\tilde{\kappa}^2))} - \frac{u \mathcal{F}_1^2[1,\frac{2+u}{2},\frac{4+u}{2},-\frac{1}{\tilde{q}_z^2+\tilde{\kappa}^2}]} {2(2+u)(\tilde{q}_z^2+\tilde{\kappa}^2)}\right] \mathcal{C}_{n,m}^{j+g;l+h} \mathcal{D}_{j+g,l+h},
\end{equation}
\end{widetext}
where $\mathcal{F}_1^2$ is the hypergeomentric function, $j,l,g,h$ denote the expansion order with respect to the in-plane momentum, $u=j+l+g+h$ is an integer, and the expressions of $\mathcal{C}$ and $\mathcal{D}$ can be found in Appendix A, where all the details about the deviation of Eq.(17) are summarized.

\subsection{Formulation of Magneto-conductivities}

As usual, the conductivities can be related to the $Q$-matrix, which can be diagrammatically expressed as Fig.~1(b). In terms of Green's function, the $Q$-matrix is given by:
\begin{align}
&Q_{ij}\left(i\omega\right) \nonumber \\
=&\frac{2e^{2}v^{2}T}{c} \sum_{m} \int \frac{dk_{y}dk_z}{4\pi^2}dx' \textrm{Tr}\left[\begin{array}{c}\sigma_{i}\hat{G}^{(0)}\left(\omega_{m}+\omega\right) \\ \sigma_{j} \hat{G}^{(0)}\left(\omega_{m}\right)\end{array}\right],
\end{align}
where $\hat{G}^{(0)}$ can be under either of the representations. Then the static conductivities can be obtained from:
\begin{equation}
\sigma_{ij}=\lim_{\omega \rightarrow 0} \frac{icQ_{ij}^{R}\left(\omega\right)}{\omega},
\end{equation}
where the superscript $R$ denotes to take the retarded part of the $Q$-matrix.

Notice that the Green's functions under two different representations are related by unitary transformations, and this difference would be removed, once we take the trace in Eq.(15). Thus, the conductivities calculated from two different representations are identical. This is actually expected, because the physical quantities (conductivities) should be independent of the representation chosen.

\subsection{chemical potential}

In this work, we are interested in a general situation where the chemical potential is not zero. In other words, there are finite charge carriers in the system. Then when we change the external magnetic field, the occupation of Landau levels changes, which thus changes the chemical potential of the system. Supposing the carrier density is $n_{0}$, the chemical potential can be determined from the following equation:
\begin{equation}
\frac{1}{\pi \ell_B^2} \left[\frac{\mu}{v} + \sum_{n=1}\int_{-\infty}^{\infty}dk_z \left(n_n^+(vk_z) + 1-n_n^-(vk_z)\right)\right] = n_{0},
\end{equation}
where $n_n^\lambda(vk_z)$ is the Fermi-Dirac distribution for the $\lambda n$ Landau level. In this way, the system of interested is just determined by the carrier density and external magnetic field.

\section{The numerical results}

For the convenience of calculations, we choose the thermal energy $T$ as the energy unit and normalize all the quantities with respect to the thermal energy $T$. Then we have a set of dimensionless normalized quantities: $\tilde{k}=vk_{z}/T$, $\tilde{\mu}=\mu/T$ and $\tilde{\beta}=2v^2/(T^2 \ell_B^2)$. We will consider both the high and low temperature cases, which is determined by the ratio between $T$ and $v/\ell_B$.

\begin{figure}[!th]
  \centering
  \includegraphics[width=0.8\columnwidth]{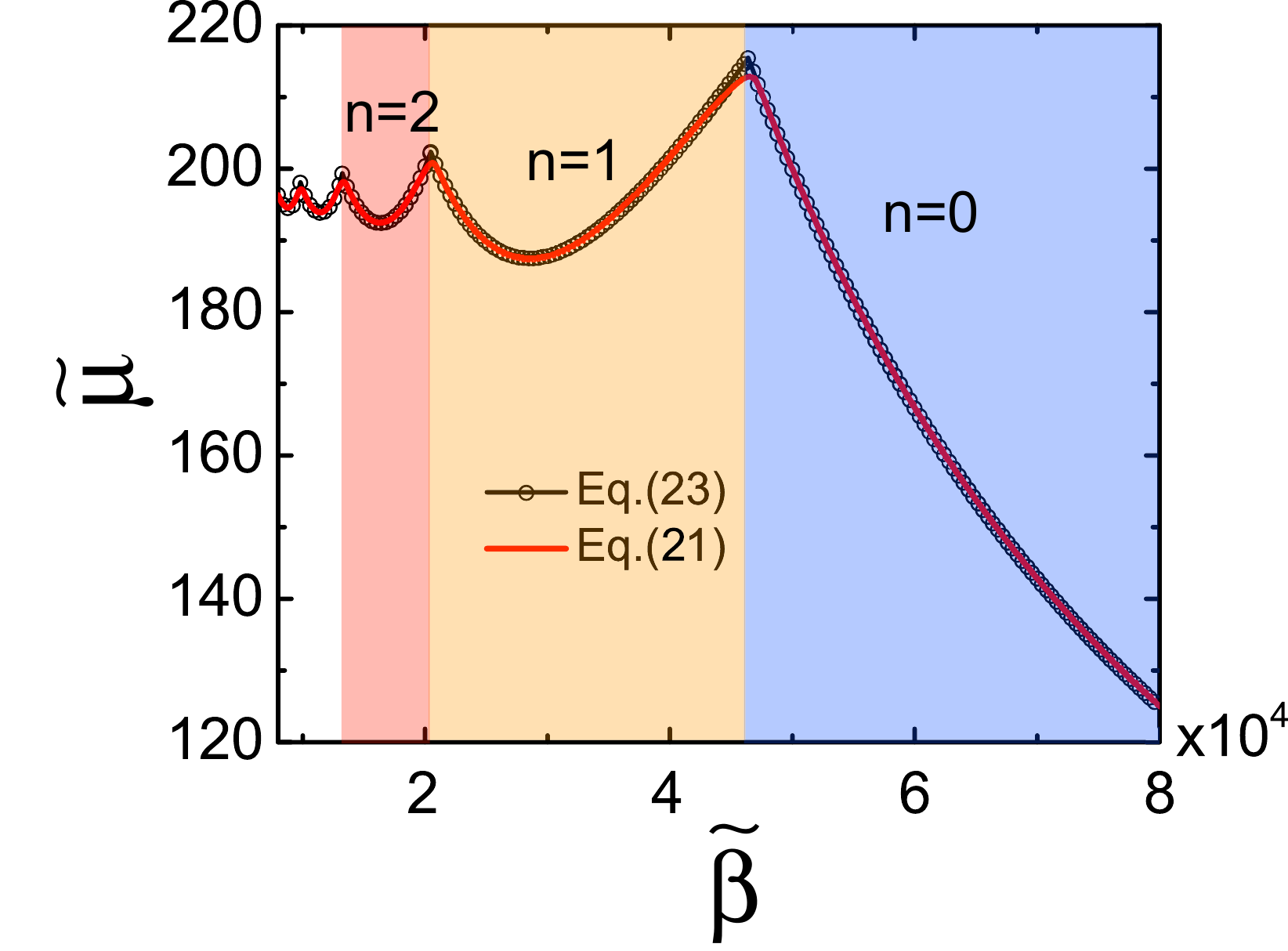}
  \caption{The normalized chemical potential $\tilde{\mu}$ as a function of the normalized magnetic field $\tilde{\beta}\sim B$. The number $n$ denotes the highest Landau level crossed by the chemical potential within the color highlighted region. The normalized carrier density is set to be $\tilde{n}_{0}=10^7$.}
   \label{fig:urtraj2}
\end{figure}

\subsection{Chemical potential and Hall conductivity}

Chemical potential and Hall conductivity is independent of the self-energy approximations. In this section we would first understand the properties of these two quantities. By using these dimensionless quantities, the equation determining the chemical potential becomes:
\begin{equation}
\tilde{\beta} \left[\tilde{\mu} + \sum_{n=1}^{N}\int_{-\infty}^{\infty} d\tilde{k}_{z} \left(n_n^+(\tilde{k}) + 1-n_n^-(\tilde{k})\right)\right]  = \tilde{n}_{0},
\end{equation}
where $\tilde{n}_{0}=\frac{2\pi v^{3}n_{0}}{T^{3}}$ and the energies appearing in $n_n^\lambda$ are scaled with respect to $T$ already. Similarly, the Hall conductivity with these dimensionless quantities can be obtained from Eq.(B1):
\begin{widetext}
\begin{equation}
\sigma_{xy} = \frac{-e^2 \tilde{\beta} T}{ 16\pi^2v} \int d\tilde{k} \sum_{\gamma,\gamma'}\sum_{n=1} \left[\tanh \frac{\tilde{E}_{n-1}^{\gamma'}\left(\tilde{k}\right)-\tilde{\mu}}{2} - \tanh \frac{\tilde{E}_{n}^{\gamma}\left(\tilde{k}\right)-\tilde{\mu}}{2} \right] \frac{\left(\chi_{n}^{-\gamma}\left(\tilde{k}\right) \right)^{2} \left(\chi_{n-1}^{\gamma'}\left(\tilde{k}\right) \right)^{2}}{\left(\tilde{E}_{n-1}^{\gamma'}\left(\tilde{k}\right)-\tilde{E}_{n}^{\gamma}\left(\tilde{k}\right)\right)^{2}} .
\end{equation}
\end{widetext}
Based on Eq.(21) and (22), we calculate the chemical potential and Hall conductivity in different parameter regimes and analyze their properties.

\begin{figure}[!th]
  \centering
  \includegraphics[width=1\columnwidth]{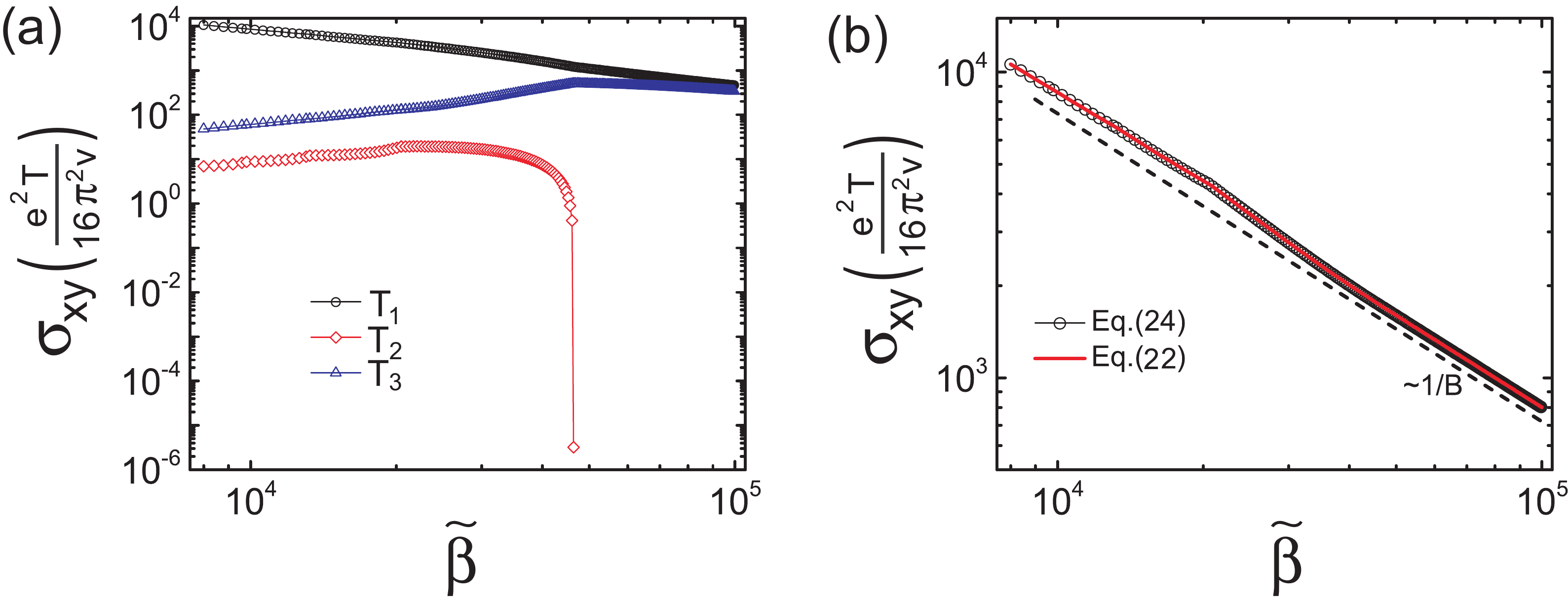}
  \caption{(a) the plot of the three terms in Eq.(24) composing the Hall conductivity $\sigma_{xy}$ as functions of $\tilde{\beta}$; (b) the Hall conductivity $\sigma_{xy}$ as a function of $\tilde{\beta}$. The red solid line is obtained numerically based on Eq.(22), while the black curve with open circles is obtained under the low temperature approximation Eq.(24). The black dashed line presents the $1/B$ scaling.}
   \label{fig:urtraj3}
\end{figure}

\subsubsection{The low temperature case $T\ll \sqrt{2v^{2}/\ell_B^2}$}

We are interested in the properties of the Landau levels with small index $n$. To make the magnetic field to be large, we assume the normalized carrier density as $\tilde{n}_{0}=10^7$. Then the chemical potential can be determined first by Eq.(21). On the other hand, since the temperature is low, the integral over the Fermi distribution functions can be evaluated by substituting the Fermi distribution by the Heaviside step function ($\frac{1} {e^{(E_{n}^{+}(\tilde{k}_{z},\tilde{\beta})\pm\tilde{\mu})}+1}\approx \Theta(E_{n}^{+}(\tilde{k}_{z},\tilde{\beta})\mp\tilde{\mu})$). Then we obtain the equation determining the chemical potential as:
\begin{equation}
\tilde{\beta} \left[\tilde{\mu} + \sum_{n=1}^{N}2\sqrt{\tilde{\mu}^2-n\tilde{\beta}}\right] =\tilde{n}_{0},
\end{equation}
where $N$ is the highest Landau level below the chemical potential.

In Fig.~2, the chemical potential determined by Eq.(23) is shown by the black curve with open circles. For the comparison, the numerical results based on Eq.(21), which has accounted the effect of temperature are presented by the red curve in the same figure. It turns out that $\tilde{\mu}$ changes with $\tilde{\beta}$ non-monotonously, and a clear increment of $\tilde{\mu}$ at the bottom of each Landau level is seen. The two methods give consistent results, except for some obvious temperature corrections at the bottom of each Landau level due to the high density of states.

Similarly, when $T$ is small in compared with $\sqrt{\tilde{\beta}}$, the Hall conductivity $\sigma_{xy}$ can be further simplified to:
\begin{equation}
\sigma_{xy} = T_1 +T_2+T_3,
\end{equation}
where:
\begin{equation}
T_1 = \frac{e^2 \tilde{\beta} T}{ 8\pi^2v} \sum_{n=1}^{N+1} b \int_{k_{n}^{\mu}}^{k_{n-1}^{\mu}} d\tilde{k} \frac{\left(\chi_{n}^{-}\left(\tilde{k}\right) \right)^{2} \left(\chi_{n-1}^{+}\left(\tilde{k}\right) \right)^{2}}{\left(\tilde{E}_{n-1}^{+}\left(\tilde{k}\right)-\tilde{E}_{n}^{+}\left(\tilde{k}\right)\right)^{2}},
\end{equation}
\begin{equation}
T_2=\frac{e^2 \tilde{\beta} T}{ 8\pi^2v} \sum_{n=2}^{N+1} \int_{k_{n}^{\mu}}^{k_{n-1}^{\mu}} d\tilde{k} \frac{\left(\chi_{n}^{-}\left(\tilde{k}\right) \right)^{2} \left(\chi_{n-1}^{-}\left(\tilde{k}\right) \right)^{2}}{\left(\tilde{E}_{n-1}^{-}\left(\tilde{k}\right)-\tilde{E}_{n}^{+}\left(\tilde{k}\right)\right)^{2}},
\end{equation}
and
\begin{equation}
T_3=\frac{e^2 \tilde{\beta} T}{ 8\pi^2v} \sum_{n=1}^{N+1} \int_{k_{n}^{\mu}}^{k_{n-1}^{\mu}} d\tilde{k} \frac{\left(\chi_{n}^{+}\left(\tilde{k}\right) \right)^{2} \left(\chi_{n-1}^{+}\left(\tilde{k}\right) \right)^{2}}{\left(\tilde{E}_{n-1}^{+}\left(\tilde{k}\right)-\tilde{E}_{n}^{-}\left(\tilde{k}\right)\right)^{2}}.
\end{equation}
In the above $N$ denotes the highest Landau level crossed by the chemical potential, $k_{n}^{\mu}$ means the Fermi wave vector for the $n$ Landau level, $b=1$ for $n=1$, and $b=2$ for other cases. The first term in Eq.(24) contains the process between positive Landau levels, while the last two terms involve the processes between the positive and negative Landau levels.

In Fig.~3(a), we plot the three parts of $\sigma_{xy}$ as functions of $\tilde{\beta}$. When only the lowest Landau level is relevant, the second term simply gets vanishing, while both the first term and the third term provide significant contribution to the Hall conductivity $\sigma_{xy}$. Based on Eq.(24), with the chemical potential under this condition, the Hall conductivity $\sigma_{xy}$ can be written as:
\begin{widetext}
\begin{equation}
\sigma_{xy}=\frac{e^2 \tilde{\beta} T}{ 8\pi^2v}\left[ \int_{0}^{k_{0}^{\mu}} d\tilde{k} \frac{\left(\chi_{1}^{-}(\tilde{k})\right)^2 \left(\chi_{0}^{+}(\tilde{k})\right)^2} {\left(\tilde{E}_{0}^{+}\left(\tilde{k}\right) -\tilde{E}_{1}^{+}\left(\tilde{k}\right)\right)^2} + \int_{0}^{k_{0}^{\mu}} d\tilde{k} \frac{\left(\chi_{1}^{+}(\tilde{k})\right)^2 \left(\chi_{0}^{+}(\tilde{k})\right)^2} {\left(\tilde{E}_{0}^{+}\left(\tilde{k}\right) -\tilde{E}_{1}^{-}\left(\tilde{k}\right)\right)^2}\right],
\end{equation}
\end{widetext}
where the first term is the approximation taken by Abrikosov \cite{Abrikosov98} containing the processes between the positive Landau levels, and the second term is from $T_3$ in Eq.(24) containing the processes between the positive and negative Landau levels. When $\tilde{k}^{2} \ll \tilde{\beta}$, $\chi_{1}^{-}(\tilde{k})\approx\chi_{1}^{+}(\tilde{k})$, and the chemical potential is very close to the neutrality point, which means $E_{1}^{+}-\mu \approx -(E_{1}^{-}-\mu)$ or particle-hole symmetry preserves approximately. Therefore, when $\tilde{k}^{2} \ll \tilde{\beta}$, the first term is equal to the second term in Eq.(28), and the Abrikosov's approximation misses a factor $2$ in the final result of $\sigma_{xy}$. Accordingly, in Fig.~3(a) with the increasing of $\tilde{\beta}$, the difference between $T_1$ and $T_3$ becomes smaller and smaller.

In Fig.~3(b), the Hall conductivity $\sigma_{xy}$ obtained by the low temperature approximation (Eq.(24)) is shown by the black curve with circle symbols. The Hall conductivity $\sigma_{xy}$ obtained by the direct numerical evaluation of Eq.(22) is shown by the red curve in the same figure for the comparison. It turns out that both methods give very good consistency. Moreover, a precise $1/B$ behavior is shown, when the condition $\mu \ll \sqrt{2v^2\beta}$, which was proposed in the seminar work by Abrikosov, is fulfilled. When $\tilde{\beta}$ is not large enough, the behavior of $\sigma_{xy}$ deviates from $1/B$ slightly.

\begin{figure}[!th]
  \centering
  \includegraphics[width=1\columnwidth]{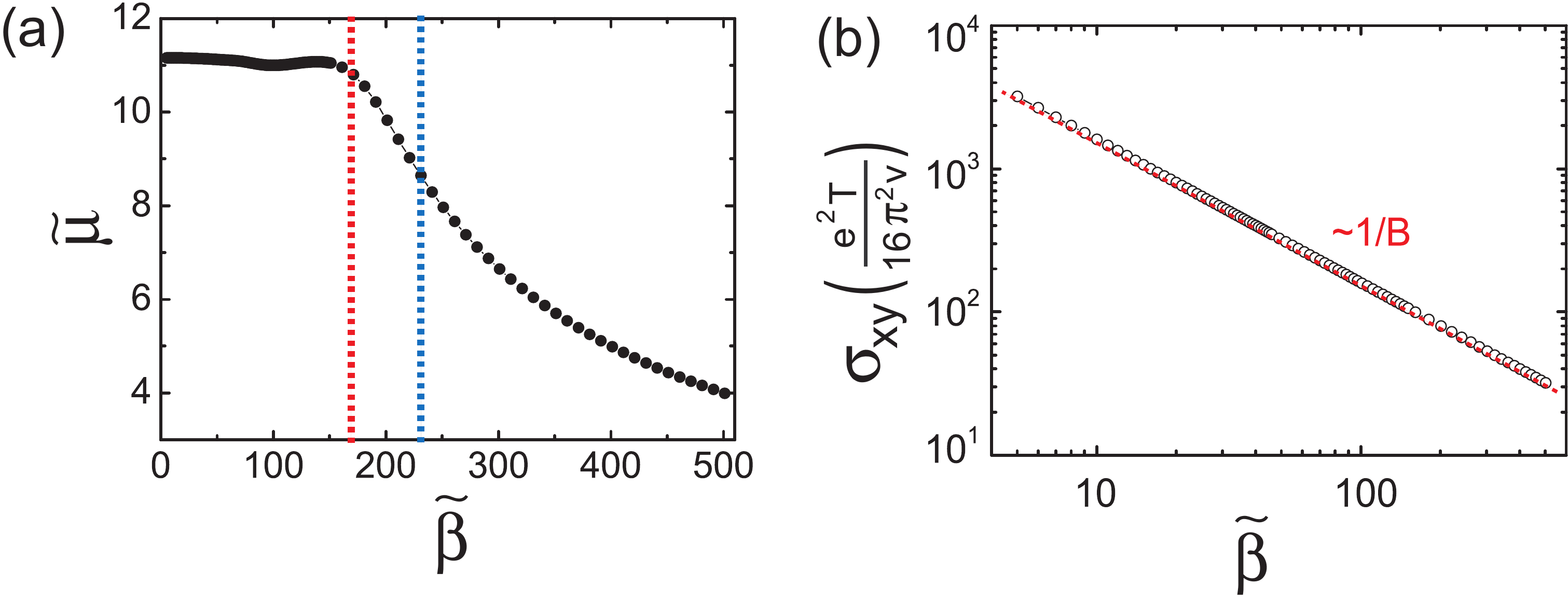}
  \caption{(a) the normalized chemical potential $\tilde{\mu}$ versus $\tilde{\beta}$: the red dashed line denotes the boundary between $n=0$ and $n=1$ Landau levels. When the magnetic field is larger than the position denoted by the blue dashed line, the $\sim1/B$ behavior of the longitudinal conductivity $\sigma_{xx}$ appears (see Fig.7). (b) the plot of the Hall conductivity $\sigma_{xy}$ as a function of $\tilde{\beta}$. We notice that within this parameter regime, the Hall conductivity follows the $~1/B$ behavior well. In the above calculations, the normalized carrier density is $\tilde{n}_{0}=2\times10^3$.}
   \label{fig:urtraj4}
\end{figure}

\subsubsection{The high temperature limit $T\sim \sqrt{2v^{2}/\ell_B^2}$}

To reduce the difficulties in numerical calculation, the normalized carrier density is set to be $\tilde{n}_{0}=2\times10^3$. With this carrier density, $\sqrt{2v^{2}/\ell_B^2}$ is about $2$ times of $T$ at the smallest magnetic field. The relation between the normalized chemical potential $\tilde{\mu}$ and $\tilde{\beta}\sim B$ can be obtained from Eq.(21) and is plotted in Fig.~4(a). With the decreasing magnetic field, the chemical potential is found to increase accordingly, except for the regime with $\tilde{\beta}\in[100,200]$, where the chemical potential shows oscillations. This is because the temperature is not high enough to smear out the effect of high density states at the Landau level bottom. We also notice that when the magnetic field is smaller than $\tilde{\beta}=100$, the oscillation due to the high density of states at the Landau level bottom is smeared out by the temperature, which is already on the same order with $\sqrt{2v^{2}/\ell_B^2}$. When the magnetic field approaches to $\tilde{\beta}=0$, the normallized chemical potential gets saturated as expected. According to Eq.(21), the low-temperature boundary between the $n=0$ Landau level and the $n=1$ Landau level is about $\tilde{\beta}\approx 160$ (see the red dashed line in Fig.~4(a)). However, according to the results to be presented in the next part (see Fig.7), the $\sim1/B$ behavior of longitudinal conductivity $\sigma_{xx}$ begins to deviate, when $\tilde{\beta}$ is smaller than the position denoted by the blue dashed line. Under the situation $T\sim \sqrt{v^{2}\beta_{0}}$, the Hall conductivity $\sigma_{xy}$ can be calculated by Eq.(22) numerically, and the relation between $\sigma_{xy}$ and $\tilde{\beta}$ is plotted in Fig.~5(b). It turns out that the Hall conductivity shows the $1/B$ behavior over the whole parameter regime. Comparing with its low-temperature behavior shown in Fig.~3(b), the temperature effect further smears out the non-$1/B$ behaviors.

\subsection{Longitudinal conductivity}

After understanding the properties of chemical potential and Hall conductivity, we consider the longitudinal conductivity $\sigma_{xx}$ in this part. To be compatible with the discussion of chemical potential and Hall conductivity, we also consider low-temperature $T\ll \sqrt{2v^{2}/\ell_B^2}$ and high-temperature limit $T\sim \sqrt{2v^{2}/\ell_B^2}$. When the temperature is low, there are only the lowest few Landau levels playing important roles. Therefore, the approximation made by Abrikosov for the overlapping integral is valid. We will follow Abrikosov's method to study the longitudinal conductivity in this limit. On the other hand, the more experimental relevant situation is when the temperature is high. In this case the conductivity is determined by many Landau levels, and we have to use the method outlined in the last section to calculate the overlapping integral accurately.

\subsubsection{low-temperature limit $T\ll \sqrt{2v^{2}/\ell_B^2}$}

Notice that in this limit the thermal effect is very small, and the approximation by Abrikosov is valid. We first calculate the self-energy, which defines the relaxation time as $\frac{1}{\tau_{n,\lambda}}=-\Im \Sigma_{n,\lambda}$. Then we can follow Abrikosov to calculate the longitudinal conductivity under the eigenstate representation (see Appendix B for details). Here we assume that the carrier density is $\tilde{n}_{0}=10^7$, under which $\tilde{\mu}$ and $\sigma_{xy}$ were studied in the section III A.

Based on the expression of the relaxation time given by Eq.(B4), with the dimensionless parameters defined at the beginning of this section we can rewrite the inverse of the relaxation time as:
\begin{widetext}
\begin{align}
\frac{1}{\tilde{\tau}_{n,\lambda}(\tilde{k})}= \frac{1}{\tau_{n}^{\lambda}((\tilde{k}))T} = \tilde{N} \sum_{\zeta=\pm1}\sum_{\gamma=\pm}\int d\tilde{k}' &\bigg\{ \frac{1}{\left(\tilde{\mu}-\tilde{E}_{n+\zeta}^{\gamma}\left(\tilde{k}'\right)\right)^{2}+1} \left[ \frac{4(\mathcal{B}_{n,\lambda}^{n+\zeta,\gamma}\left(\tilde{k},\tilde{k}'\right))^{2}}{\tilde{\beta}^{2}} +(\mathcal{A}_{n,\lambda}^{n+\zeta,\gamma}\left(\tilde{k},\tilde{k}'\right))^{2}\right] \nonumber \\
&\times \left(\ln\frac{\tilde{\beta}/2+\tilde{K}^{2}(\tilde{k},\tilde{k}')}{\tilde{K}^{2}(\tilde{k},\tilde{k}')} + \frac{\tilde{K}^{2}(\tilde{k},\tilde{k}')}{\tilde{\beta}/2+\tilde{K}^{2}(\tilde{k},\tilde{k}')}-1\right)\bigg\},
\end{align}
\end{widetext}
where $\tilde{N}=v^3 N_i/(\varepsilon_{\infty}^2 T^3)$ is the dimensionless parameter describing the density of impurities, $\tilde{K}(\tilde{k},\tilde{k}')=v^2/T^2 \left[(k_z-k'_z)^2+\kappa^2\right]$, and the expressions of the parameters $\mathcal{A}_{n,\lambda}^{m,\gamma}\left(\tilde{k},\tilde{k}'\right)$ and $\mathcal{B}_{n,\lambda}^{m,\gamma}\left(\tilde{k},\tilde{k}'\right)$ are given by Eq.(B5) and Eq.(B6) in the Appendix B.

By using the dimensionless parameters defined at the beginning of this section, we can also rewrite the expression of longitudinal conductivity given by Eq.(B2) and (B3) in Appendix B into the following form:
\begin{equation}
\sigma_{xx} = \frac{e^{2}T}{v} \frac{\tilde{\beta}}{2\pi} \int \frac{d\tilde{\omega}}{2\pi} \frac{d\tilde{k}}{2\pi} \cosh^{-2}\frac{\tilde{\omega}}{2} \tilde{F}\left(\tilde{\omega},\tilde{k}\right),
\end{equation}
where
\begin{widetext}
\begin{equation}
\tilde{F}\left(\tilde{\omega},\tilde{k}\right)=\frac{1}{4} \sum_{n,\gamma,\gamma'} \frac{(\chi_{n}^{-\gamma}(\tilde{k}))^{2}/\tilde{\tau}_{n}^{\gamma}(\tilde{k})}{\left(\tilde{\omega}+\tilde{\mu} -\tilde{E}_{n}^{\gamma}\left(\tilde{k}\right)\right)^{2}+\frac{1}{(\tilde{\tau}_{n}^{\gamma}(\tilde{k}))^2}} \frac{(\chi_{n-1}^{\gamma'}(\tilde{k}))^{2}/\tilde{\tau}_{n-1}^{\gamma'}(\tilde{k})}{\left(\tilde{\omega}+\tilde{\mu} -\tilde{E}_{n-1}^{\gamma'}\left(\tilde{k}\right)\right)^{2}+\frac{1}{(\tilde{\tau}_{n-1}^{\gamma'}(\tilde{k}))^2}}.
\end{equation}
\end{widetext}

\begin{figure}[!th]
  \centering
  \includegraphics[width=0.8\columnwidth]{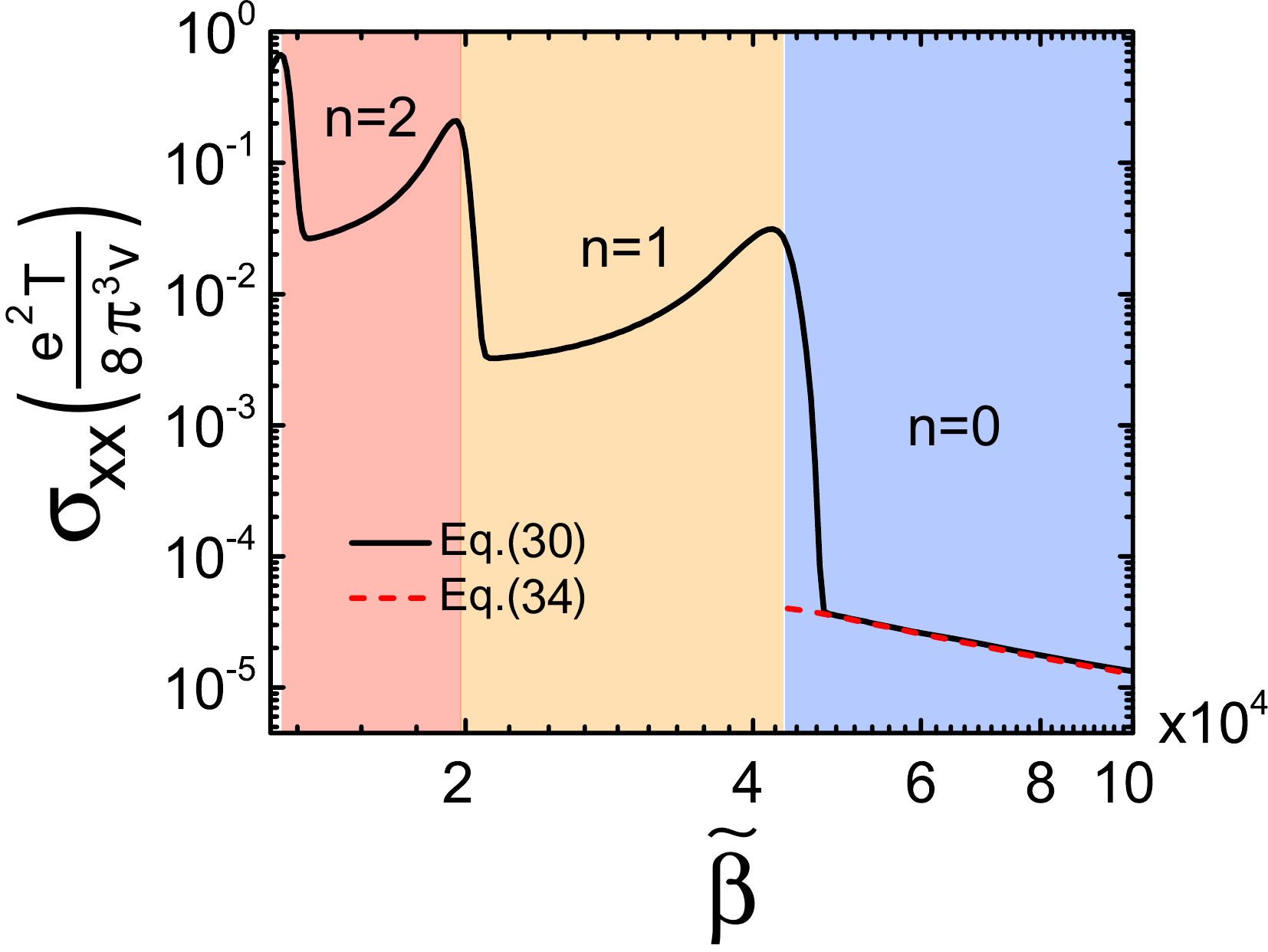}
  \caption{The plot of the calculated longitudinal conductivity $\sigma_{xx}$ versus the reduced magnetic field $\tilde{\beta}$. The number $n$ denotes the highest Landau level crossed by the chemical potential within the color highlighted region.}
   \label{fig:urtraj5}
\end{figure}

In Fig.~5, the longitudinal conductivity $\sigma_{xx}$ obtained numerically from Eq.(30) is plotted as a function of $\tilde{\beta}$. The conductivity shows a non-monotonous behavior, and at around the bottom of each Landau level a maximum in conductivity appears due to the high density of states.

When the temperature is low and the chemical potential crosses only the lowest Landau level, the expression of $\sigma_{xx}$ can be simplified from Eq.(30):
\begin{align}
\sigma_{xx} &\approx \frac{e^2T\tilde{\beta}}{4\pi^2 v} \frac{(\chi_{1}^{-}(\tilde{k}_{0}^{\tilde{\mu}}))^2 /\tilde{\tau}_{1}^{+}(\tilde{k}_{0}^{\tilde{\mu}})} {(\tilde{E}_{1}^{+}(\tilde{k}_{0}^{\tilde{\mu}})-\tilde{E}_{0}^{+}(\tilde{k}_{0}^{\tilde{\mu}}))^2} \nonumber \\
&+ \frac{e^2T\tilde{\beta}}{4\pi^2 v}  \frac{(\chi_{1}^{+}(\tilde{k}_{0}^{\tilde{\mu}}))^2 /\tilde{\tau}_{1}^{-}(\tilde{k}_{0}^{\tilde{\mu}})} {(\tilde{E}_{1}^{-}(\tilde{k}_{0}^{\tilde{\mu}})-\tilde{E}_{0}^{+}(\tilde{k}_{0}^{\tilde{\mu}}))^2},
\end{align}
where $\tilde{k}_{0}^{\mu}$ is the Fermi energy for the $n=0$ Landau level, and $\tilde{\tau}_{1}^{\pm}$ is the reduced scattering time for the $n=\pm1$ Landau level, which can be obtained from Eq.(29) under the same approximations:
\begin{align}
\frac{1}{\tilde{\tau}_{1}^{\pm}(\tilde{k}_{0}^{\tilde{\mu}})} \approx \frac{\pi \tilde{N}}{\tilde{\beta}} (\chi_{1}^{\pm}(\tilde{k}_{0}^{\tilde{\mu}}))^2 \left(\ln\frac{\tilde{\beta}+\tilde{\kappa}^{2}}{\tilde{\kappa}^{2}} + \frac{\tilde{\kappa}^{2}}{\tilde{\beta}+\tilde{\kappa}^{2}}-1\right),
\end{align}
where $\tilde{\kappa}=v\kappa/T$. Combining Eq.(32) and Eq.(33), we arrive at the expression of longitudinal conductivity under the conditions mentioned above:
\begin{equation}
\sigma_{xx} \approx \frac{e^2T\tilde{N}}{8\pi v} \frac{1}{\tilde{\beta}} \frac{4(\tilde{k}_{0}^{\tilde{\mu}})^2+2\tilde{\beta}}{(\tilde{k}_{0}^{\tilde{\mu}})^2+\tilde{\beta}} \left(\ln\frac{\tilde{\beta}+\tilde{\kappa}^{2}}{\tilde{\kappa}^{2}} + \frac{\tilde{\kappa}^{2}}{\tilde{\beta}+\tilde{\kappa}^{2}}-1\right).
\end{equation}
It is easy to check that when $\mu \ll \sqrt{2v^2\beta}$, the longitudinal conductivity $\sigma_{xx}$ is proportional to $1/\tilde{\beta}\propto 1/B$. Together with the $1/B$ behavior of $\sigma_{xy}$, the linear magnetoresistance is expected under these conditions, which is consistent with the observation of Abrikosov \cite{Abrikosov98}. To support the correctness of the above analysis, we focus on the $n=0$ case and plot $\sigma_{xx}$ obtained by Eq.(34) in the red dashed line in the Fig.~5. It turns out that the comparison with the numerical result based on Eq.(30) is very good.

\subsubsection{The high temperature limit $T\sim \sqrt{2v^{2}/\ell_B^2}$}

When the temperature is high, the charge carriers from the higher Landau levels can always play roles due to the non-negligible thermal energy. Importantly, this is the usual situation for the experiments. Therefore, to calculate the longitudinal conductivity correctly, we have to use the scheme outlined in the section II B to evaluate the self-energy to the desired accuracy within the self-consistent Born approximation.

In the above discussion, we claimed many times that the approximations made by Arikosov \cite{Abrikosov98} and Klier \cite{Klier15} are inaccurate, when the charge carriers from the higher Landau levels play important roles. Here we first demonstrate this point numerically.

\begin{figure}[!th]
  \centering
  \includegraphics[width=1\columnwidth]{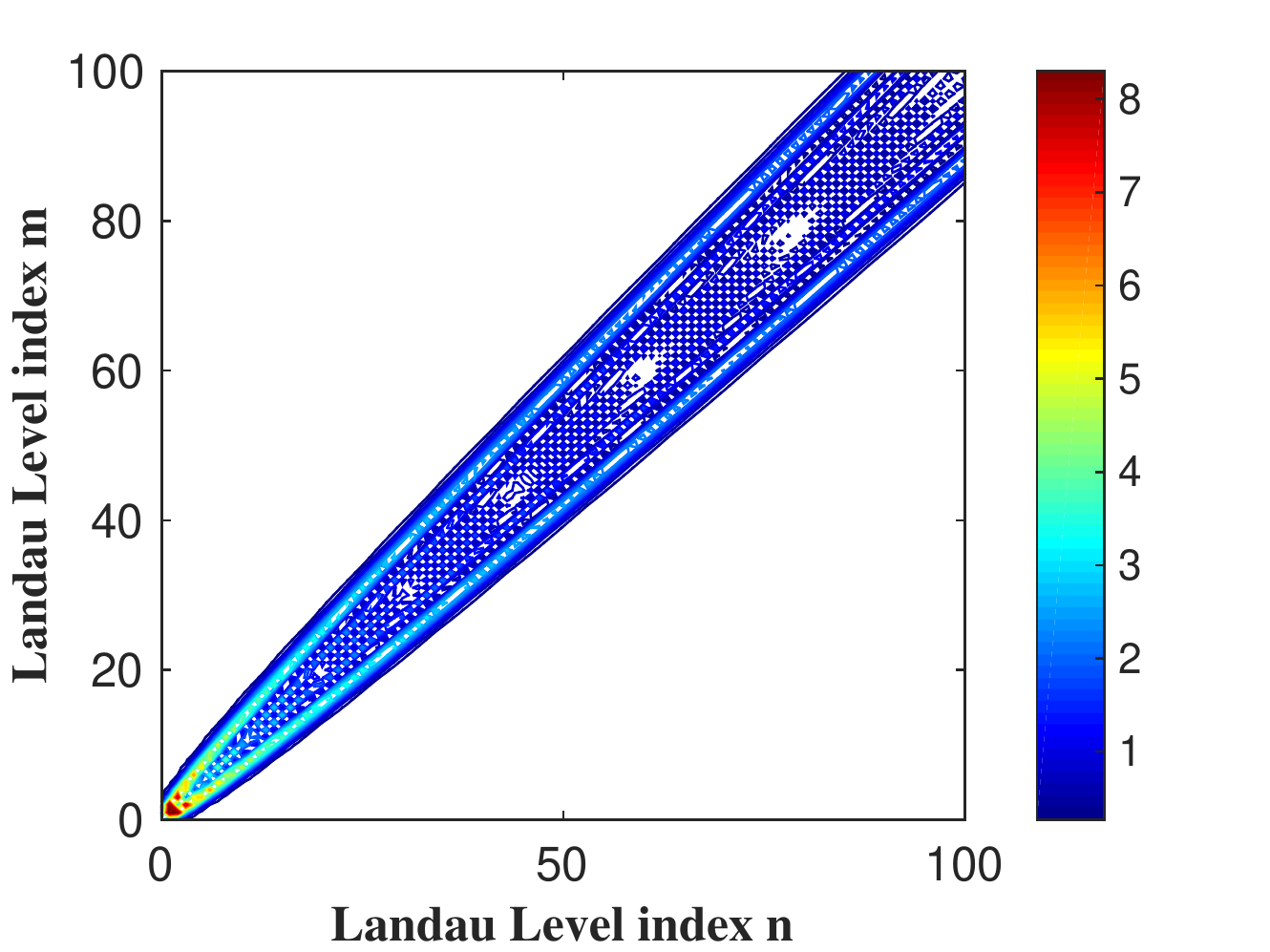}
  \caption{The plot of correlation matrix $I_{n,m}$ between Landau levels. The position with white color means that the correlation between the Landau level is $0$. In our calculation the upper limit of the expansion order is set to be $N=50$. Higher order terms with $j,g,h,l$ larger than $N=50$ are also tested, and they do not have much influence on the result above.}
   \label{fig:urtraj6}
\end{figure}

The value of the overlapping integral Eq.(17) is largely determined by the value of
\begin{equation}
I_{n,m} = \sum_{j,l,g,h} \mathcal{C}_{n,m}^{j+g;l+h} \mathcal{D}_{j+g;l+h},
\end{equation}
given the fact that the hypergeometric function slowly varies with $u=j+g+l+h$ and the other terms in Eq.(17) are independent of the Landau level index. We explicitly evaluate the correlation matrix $I_{n,m}$ between different Landau levels, and the results are shown in Fig.6. Indeed, we find that when the Landau level index is smaller, the correlation is really short-ranged, meaning that the correlation with $|m-n|>1$ can be assumed to be $0$. This is actually the approximation made by Abrikosov \cite{Abrikosov98}. However, with the increment of the Landau level index, the range of non-vanishing correlations increases. Therefore, the approximation by Abrikosov would break down under this situations. Moreover, we notice that the correlation matrix is not independent of the Landau level index, so the approximation by Klier \textit{et al} also breaks down.

The results shown in Fig.6 thus demonstrate the advantages of the present method in the calculation of self-energy. With the help of this, we calculate the longitudinal conductivity with the self-energy obtained from self-consistent Born approximation introduced in section II B. In terms of the dimensionless parameters, the longitudinal conductivity $\sigma_{xx}$ can be written as:
\begin{widetext}
\begin{equation}
\sigma_{xx} = \frac{e^2 \tilde{\beta}^2 T}{4\pi v} \int \frac{d\tilde{\omega}}{2\pi} \cosh^{-2} \left(\frac{\tilde{\omega}}{2}\right) \sum_{n} \int \frac{d\tilde{k}_z}{2\pi} \textrm{Im} \tilde{G}_{11}^{L}(\tilde{\omega},n,\tilde{k}_z) \textrm{Im} \tilde{G}_{22}^{L}(\tilde{\omega},n+1,\tilde{k}_z),
\end{equation}
\end{widetext}
where $G_{nm}^{L}$ is to denote the $(m,n)$ component of the Green's function under the Landau level representation. Here we choose the Landau level representation, because it is much convenient for the performance of self-consistent calculations. In the calculations below, we assume that charge carrier density is $\tilde{n}_0=2\times10^3$, which is also used in the last part to study the chemical potential and Hall conductivity in the high-temperature limit.

\begin{figure}[!th]
  \centering
  \includegraphics[width=0.8\columnwidth]{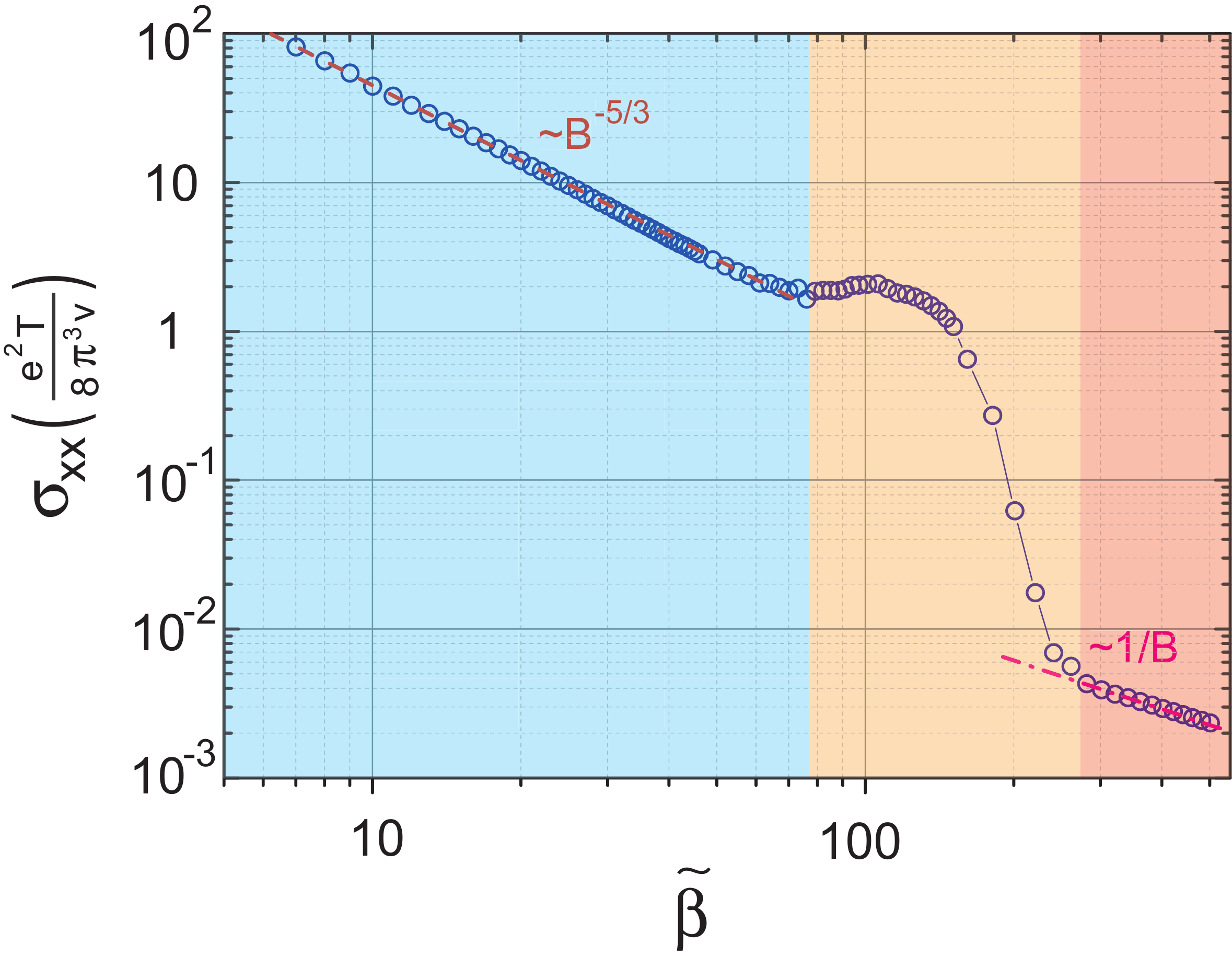}
  \caption{The plot of longitudinal magneto-conductivity $\sigma_{xx}$ as a function of $\tilde{\beta}$. The behavior of the conductivity can be divided into three regimes: 1. the large field (the red shielding region), where $\sigma_{xx}$ shows the $1/B$ behaviors, which is the same with that obtained in the low-temperature limit; 2. the middle regime (the orange shielding region), where $\sigma_{xx}$ shows the non-monotonic behavior; 3. the low field regime (the green shielding region), where $\sigma_{xx}$ shows the $B^{-5/3}$ behavior. The calculation involves the lowest $100$ Landau levels.}
   \label{fig:urtraj7}
\end{figure}

In Fig.7 we plot the longitudinal conductivity $\sigma_{xx}$ as a function of the magnetic field $\tilde{\beta}$. Obviously, the behavior of the longitudinal conductivity $\sigma_{xx}$ can be divided into three regimes. When the magnetic field is large, the longitudinal conductivity shows the $\sim1/B$ behavior (the red shielding region in Fig.7), because the charge carriers are mainly from the zeroth Landau level. When the magnetic field is smaller but still much larger than the thermal energy $T$, the non-monotonic behavior (see the orange shielding region in Fig.7) similar with the low temperature limit (see Fig.5) can be seen. This is due to the fact that the small thermal energy does not yet suppress the effect caused by the high density of states at the Landau level bottoms. Interestingly, the longitudinal conductivity $\sigma_{xx}$ shows the $\sim B^{-5/3}$ behavior in the low-field regime (see the blue shielding region in Fig.7). This universal $\sim B^{-5/3}$ behavior of $\sigma_{xx}$ is a very important feature of self-energy approximation in the high-temperature limit.

\begin{figure}[!th]
  \centering
  \includegraphics[width=1\columnwidth]{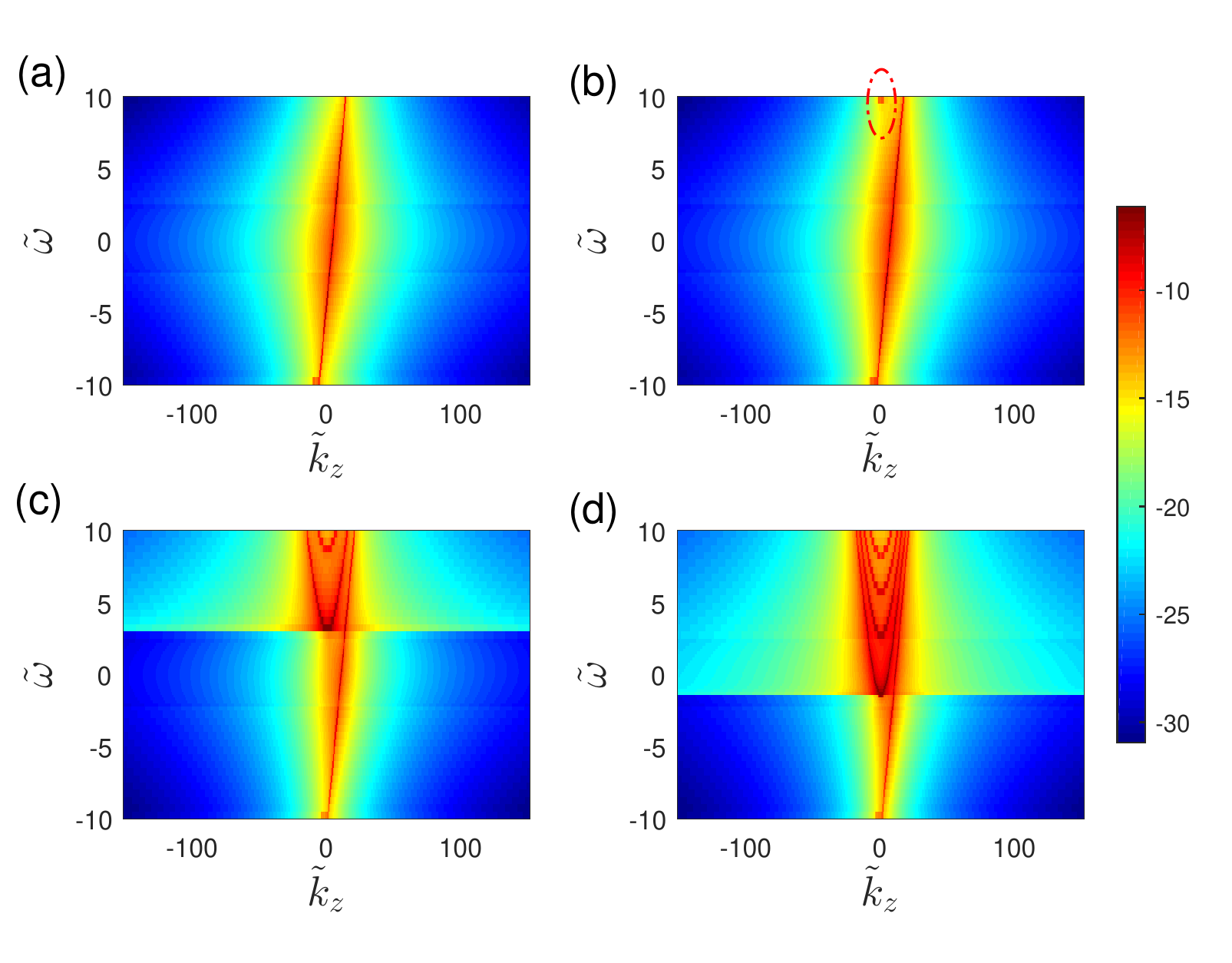}
  \caption{The log plot of the integrand of Eq.(36): (a) for $\tilde{\beta}=480$; (b) for $\tilde{\beta}=280$; (c) for $\tilde{\beta}=180$; (d) for $\tilde{\beta}=90$. The red dashed-dotted circle in (b) is to highlight the resonance of the $n=1$ Landau level. This indicates that the $n=1$ Landau level begins to contribute to $\sigma_{xx}$.}
   \label{fig:urtraj8}
\end{figure}

To further understand the properties of $\sigma_{xx}$, we plot the integrand of Eq.(36) at some typical values of magnetic field in Fig.8. When the magnetic field is high, there is only one branch of singularities (the red curves in Fig.8(a)) contributing to $\sigma_{xx}$, which is due to the zeroth ($n=0$) Landau level. Decreasing the magnetic field to $\tilde{\beta}=280$, the $n=1$ Landau level begins to contribute to $\sigma_{xx}$ (see Fig.8(b)). However, we have to note that the deviation from the $1/B$ behavior does not immediately happens at this magnetic field. The further reduction of magnetic field would make more and more Landau level to be involved. One can expect that when the magnetic field is very small, the singularities of the integrand would form a Weyl node shape.

Due to the chiral property of the zeroth Landau level, there is only one branch of singularities for it. However, for the higher Landau levels, there are two branches. Therefore, when the $n=1$ Landau level begins to play non-negligible roles, the extra branch of singularity breaks the total chiral property and provides a dramatic jump of $\sigma_{xx}$ around $\tilde{\beta}=200$, which is another important feature of the self-energy approximation applied for Weyl semimetal.

\subsubsection{Comparison with experimental data}

Before we close the section, let us compare our results with an available experimental data. We focus on the experimental relevant high-temperature limit. From the Hall and longitudinal conductivity obtained in the last parts, we can extract the magnetoresistance:
\begin{equation}
\rho_{xx} = \frac{\sigma_{xx}}{\sigma_{xx}^2+\sigma_{xy}^2}.
\end{equation}
The calcaulated magnetoresistance $\rho_{xx}$ is shown in Fig.9(a). When the field is low, we notice that $\sigma_{xy}\gg\sigma_{xx}$, so the magnetoresistance shows the $B^{1/3}$ behavior. In the middle-field regime, due the to small thermal energy, the non-monotonic behavior of $\rho_{xx}$ is caused by the high density of states at the Landau level edges. In the high-field regime, we find the linear magnetoresistance.

\begin{figure}[!th]
  \centering
  \includegraphics[width=0.8\columnwidth]{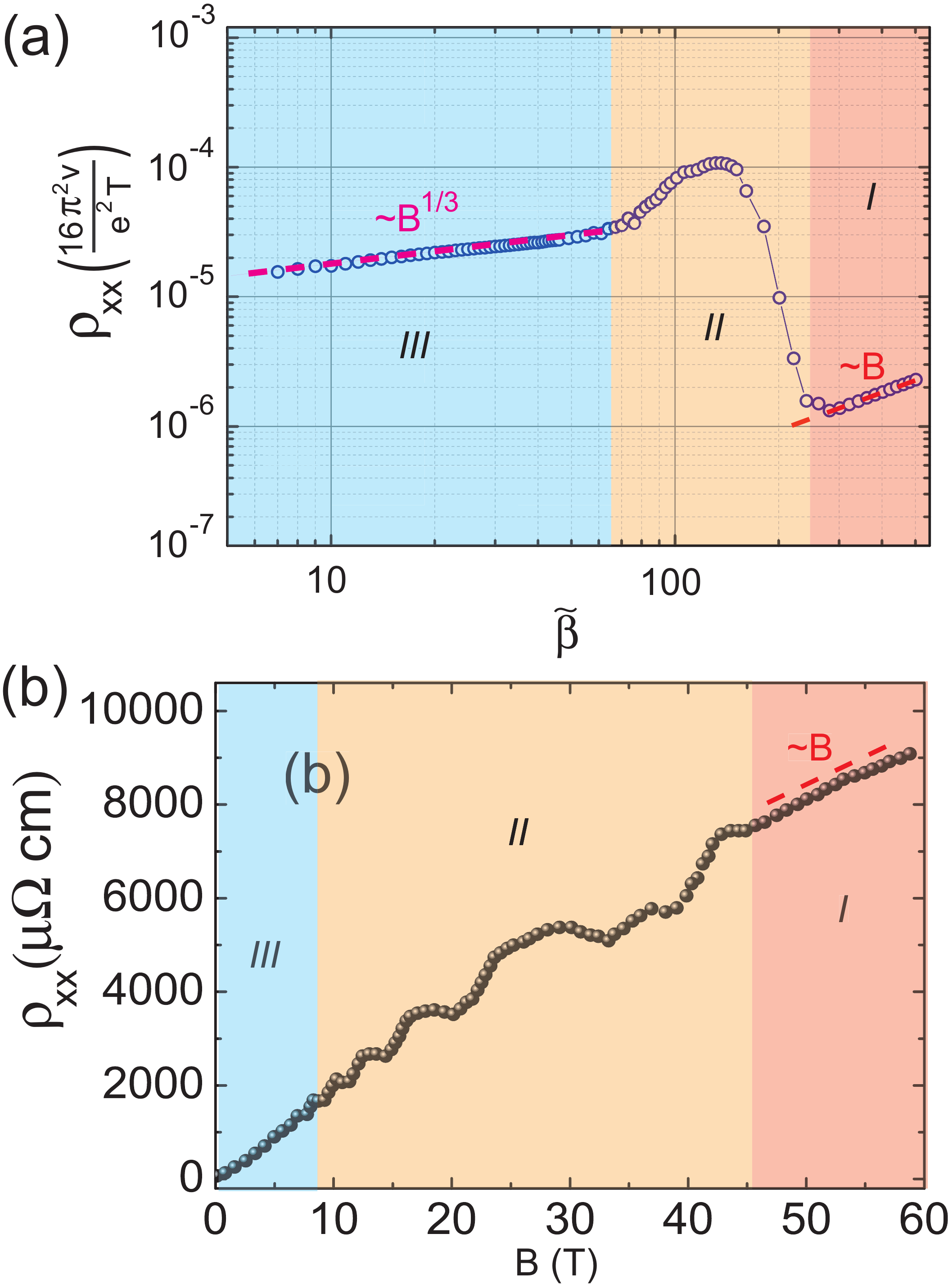}
  \caption{The comparison of magnetoresistance: (a) the magnetoresistance $\rho_{xx}$ obtained from the numerical calculations as a function of magnetic field; (b) the experimental measured magnetoresistance $\rho_{xx}$ as a function of magnetic field \cite{Zhao15}. Both data can be divided into three regimes: the low-field power law regime, the middle-field oscillation regime, and the high-field $\sim B$ regime.}
   \label{fig:urtraj9}
\end{figure}

For the comparison, the experimental measured magnetoresistance is shown in Fig.9(b). The main difference between the results obtained from the self-energy approximation and those from measurements are: 1. in the low field the exponents of power law from the experiment data is larger than that from our calculation; 2. in the middle field regime, the oscillation amplitude from our calculation is too large in comparison with that of experimental data.

\section{Discussion and conclusion}

In this work, we studied the magneto-transport based on a low-energy effective model of a single Weyl node within the self-energy approximation for charged impurities. Actually, the Weyl nodes should appear in pairs. However, as long as we focus on the bulk and is not interested in the interplay between different nodes, a single node model is enough to obtain reasonable results. Under this situation, the effect of multi-node can be accounted simply by multiplying the results of single node model by the number of nodes. The results present here are valid, when the sample is clean with $\tau \ll \sqrt{v^2\beta_0}$ and the smearing of Landau level caused by disorder can be ignored.

In the low-temperature limit ($T \ll \sqrt{v^2\beta_0}$), within the self-energy approximation, the chemical potential varies with the change of external magnetic field non-monotonously. In this limit, the Hall conductivity basically follows the $1/B$ behavior except for a small derivation when chemical potential crosses a Landau level band bottom. For the longitudinal magneto-conductivity, the following magnetic field dependence appears: it shows $1/B$ behavior, if the charge carriers are contributed mainly by the $n=0$ Landau level; On the other hand, the longitudinal magneto-conductivity shows the `step-like' structure and is with steep increasing at the edges of Landau levels due to the high density of states, when the the chemical potential locates at $n\geq 1$ Landau levels.

When the temperature is comparable with the gap between Landau levels ($T \sim \sqrt{v^2\beta_0}$), the chemical potential increases with the decreasing external magnetic field and gradually saturates to the $\tilde{\beta}=0$ value, and the Hall conductivity within this parameter regime shows nicely the $1/B$ behavior. In this parameter regime, the magnetic field dependence behavior of longitudinal magneto-conductivity can be divided into three regimes. The same with the low temperature case, the $1/B$ behavior appears, when the charge carriers are mainly from the $n=0$ Landau level. Therefore, within the self-energy approximation, no matter the temperature is high or low, the linear magneto-resistance can appear robustly. In the middle field regime $\sigma_{xx}$ shows non-monotonic behavior, while in the low-field regime the longitudinal conductivity shows a $~B^{-5/3}$ behavior with the change of the external magnetic field. Consequently, in the low-field regime, the self-energy approximation predicts the $~B^{1/3}$ behavior for the magnetoresistance.

The major failure of self-energy approximation is that it predicts a dramatic change of $\sigma_{xx}$, when the $n=1$ Landau level begins to play roles. Consequently, this leads to the non-monotonic change of the magnetoresistance with the external magnetic field and a relatively large oscillation amplitude in comparison with the experiment data \cite{Zhao15}. Moreover, in experiments the magnetoresistance appears to be almost linear and associated with quantum oscillations \cite{Liang15,Zhao15}, which indicates that the linear magneto-resistance can appear even when the chemical potential intersects with higher Landau levels. However, the self-energy approximation predicts a $B^{1/3}$ scaling of the magnetoresistance in this parameter regime. We conclude that the self-energy approximation of this model does not explain the linear magnetoresistance appeared at the low field in experiments. The almost linear magnetoresistance associated with quantum oscillations observed in the experiments really calls for a quantum approach to understand. Given the success of the semi-classical picture of guiding center motion in describing the linear magnetoresistance in Weyl semi-metal \cite{Song15}, it is very interesting to see whether it can be extended to a quantum version so that we can understand better about the associated quantum oscillations.

\begin{acknowledgments}
PAL would like to acknowledge the support of DOE under grant DE-FG01-03-ER46076. He also thank the hospitality of the Institute of Advanced studies at the Hong Kong University of Science and Technology. KTL thanks the support of HKRGC and Croucher Foundation through HKUST3/CRF/13G, 602813, 605512, 16303014 and Croucher Innovation Grant.
\end{acknowledgments}

\appendix
\section{The scheme to evaluate the overlapping integrals}

In this appendix, we outline the general scheme to calculate the overlapping integrals. For the clarity, we just use the Landau level representation, and this procedures can be also applied to eigenstate representation.

Let us consider impurities in real space described by the following scattering potential:
\begin{equation}
V_{imp}(\vec{r}) = \sum_{i} U(\vec{r}-\vec{r}_i),
\end{equation}
where $\vec{r}_i$ denotes the position of an impurity in space. From the second order correction to the bare Green's function, we can extract the form of self-energy, which can be generally written as:
\begin{equation}
\Sigma_\nu(i\omega) = \sum_{\chi} \left|\left( \int d^3r V_{imp}(\vec{r}) \varphi_\nu^{\dag}(\vec{r}) \varphi_\chi(\vec{r})  \right)\right|^2 G_\chi^{(0)} (i\omega_n),
\end{equation}
in real space.
For our present case, we have to first write down the impurity scattering potential in the Fourier space, then insert real space representation to the Green's function defined in Eq.(6) or (10), and finally integrate out $d^3r$ and $d^3r'$ to find that:
\begin{widetext}
\begin{equation}
\Sigma_n(i\omega_n) = N_{imp} \int \frac{d^3q}{(2\pi)^3} \sum_m |U(\vec{q})|^2 \left|\int dx \Phi_n^{\dag}(x-k_y\ell_B^2) e^{iq_xx}\Phi_m(x-(k_y+q_y)\ell_B^2)\right|^2 \tilde{G}_m^{(0)}(i\omega_n,k_z+q_z),
\end{equation}
\end{widetext}
which can be rearranged into Eq.(13) in the main text above. In what follows, we explicitly use the Landau level representation. The same procedure can be also applied to the eigenstate representation. Then, the overlapping integral is expressed as:
\begin{equation}
\mathcal{I}_{n,m} = \frac{1}{8\pi^2\ell_B^2} \int_{|\tilde{q}|\leq1} d\tilde{q}^2 |U(\tilde{\vec{q}})|^2 \left| \mathcal{J}_{n,m} + \mathcal{J}_{n-1,m-1} \right|^2,
\end{equation}
where:
\begin{equation}
\mathcal{J}_{n,m}(\tilde{q}_x,\tilde{q}_y)=\int d\tilde{x} \phi_n(\tilde{x}) e^{i\tilde{q}_x \tilde{x}} \phi_m(\tilde{x}-\tilde{q}_y).
\end{equation}
In the above, $\tilde{x}=(x-k_y\ell_B^2)/\ell_B$ and $\tilde{q}_{x,y}=q_{x,y}\ell_B$ are dimensionless parameters respect to the magnetic length $\ell_B$.

Given the fact that $|\tilde{q}|\leq1$, we can expand the overlapping integral with respect to the in-plane momentum, and this gives:
\begin{equation}
\mathcal{J}_{n,m}(\tilde{q}_x,\tilde{q}_y) = \sum_{j,l} \frac{(i\tilde{q}_x)^j}{j!} \frac{(-\tilde{q}_y)^l}{l!} \int d\tilde{x} \phi_n(\tilde{x}) \tilde{x}^j \partial_{\tilde{x}}^l \phi_m(\tilde{x}).
\end{equation}
This equation can be numerically tested, and the right-hand side is almost equal to the left, once the expansion reaches high enough order, which depends on the index of Landau level $n$ and $m$.

The integral in Eq.(A6) can be evaluated by using the ladder operators. For example, the position and momentum operator can be written by the ladder operators:
\begin{equation}
\begin{cases}
\tilde{x} = \frac{1}{\sqrt{2}} \left( \hat{a}^{\dag}+\hat{a} \right), \\
\partial_{\tilde{x}} = \frac{1}{\sqrt{2}} \left( \hat{a}-\hat{a}^{\dag}  \right).
\end{cases}
\end{equation}
Under the Landau level representation, the ladder operator can be expressed as the following matrix form:
\begin{equation}
\hat{a}^{\dag} = \left(\begin{array}{ccccc} 0 & 0 & 0 & 0 & \cdots \\ 1 & 0 & 0 & 0 & \ddots \\ 0 & \sqrt{2} & 0 & 0 & \ddots \\ 0 & 0 & \sqrt{3} & 0 & \ddots \\ \vdots & \ddots & \ddots & \ddots & \ddots \end{array}\right) = \mathcal{A}_{+},
\end{equation}
and
\begin{equation}
\hat{a} = \left(\begin{array}{ccccc} 0 & 1 & 0 & 0 & \cdots \\ 0 & 0 & \sqrt{2} & 0 & \ddots \\ 0 & 0 & 0 & \sqrt{3} & \ddots \\ 0 & 0 & 0 & 0 & \ddots \\ \vdots & \ddots & \ddots & \ddots & \ddots \end{array}\right) = \mathcal{A}_{-}.
\end{equation}
The dimension of the matrix would be determined by the accuracy. This matrix form is very useful and reduces the integral becomes an element of the product of matrix following the identities below:
\begin{equation}
\begin{cases}
\int d\tilde{x} \phi_n(\tilde{x}) \tilde{x}^j \phi_m(\tilde{x}) = \left(\frac{1}{\sqrt{2}}\right)^j \left[(\mathcal{A}_+ + \mathcal{A}_-)^j\right]_{n+1,m+1}, \\
\int d\tilde{x} \phi_n(\tilde{x}) \partial_{\tilde{x}}^j \phi_m(\tilde{x}) = \left(\frac{1}{\sqrt{2}}\right)^j \left[(\mathcal{A}_- - \mathcal{A}_+)^j\right]_{n+1,m+1},
\end{cases}
\end{equation}
where the subscript $n+1$ means the $n+1$th row of the matrix, and $m+1$ indicates the $m+1$th column of the matrix.  This labeling is to account that there is the zeroth Landau level.Then it is convenient to define:
\begin{equation}
J_{n,m}^{j,l} = \left(\frac{1}{\sqrt{2}}\right)^{j+l} \left[(\mathcal{A}_+ + \mathcal{A}_-)^j (\mathcal{A}_- - \mathcal{A}_+)^l\right]_{n+1,m+1}.
\end{equation}
Therefore, with these elements we can write:
\begin{equation}
\left|\mathcal{J}_{n,m}+\mathcal{J}_{n-1,m-1}\right|^2
= \sum_{j,l,g,h} \mathcal{C}_{n,m}^{j+g,l+h} \tilde{q}_x^{j+g} \tilde{q}_y^{l+h},
\end{equation}
where
\begin{align}
\mathcal{C}_{n,m}^{j+g,l+h} &= \frac{(i)^{j+g}(-1)^{l+h+g}}{j!l!g!h!} \nonumber \\
 &\times\left( J_{n,m}^{j,l} + J_{n-1,m-1}^{j,l} \right) \left( J_{n,m}^{g,h} + J_{n-1,m-1}^{g,h} \right),
\end{align}
The expression above has a great advantage, because except for $\tilde{q}_x^{j+g} \tilde{q}_y^{l+h}$ the others are just coefficients and independent of momentum. Insert $\tilde{q}_x^{j+g} \tilde{q}_y^{l+h}$ into the integral over $\tilde{q}_x$ and $\tilde{q}_y$,
\begin{align}
&\int_{\tilde{q}\leq1} d\tilde{q}_x d\tilde{q}_y |U(\tilde{q})|^2 \tilde{q}_x^{j+g} \tilde{q}_y^{l+h}
 \nonumber \\
= &\int_{0}^1 d\tilde{q} |U(\tilde{q})|^2 \tilde{q}^{j+g+l+h+1} \int_0^{2\pi} d\theta \cos^{j+g}(\theta) \sin^{l+h}(\theta).
\end{align}
We denote the angle integral by $\mathcal{D}_{j+g,l+h}$
\begin{equation}
\mathcal{D}_{j+g,l+h}=\int_0^{2\pi} d\theta \cos^{j+g}(\theta) \sin^{l+h}(\theta).
\end{equation}
Then take the charged impurity scattering potential as:
\begin{equation}
U(\tilde{q},\tilde{q}_z) = \frac{4\pi e^2 \ell_B^2}{\varepsilon_{\infty}} \frac{1}{\tilde{q}^2 + \tilde{q}_z^2 +\tilde{\kappa}^2},
\end{equation}
and we find that the radial integral gives:
\begin{align}
&\int_{0}^1 d\tilde{q} |U(\tilde{q})|^2 \tilde{q}^{u+1} \nonumber \\
=& \frac{8 \pi^2 e^4 \ell_B^4}{\varepsilon_{\infty}^2(\tilde{q}_z^2+\tilde{\kappa}^2)} \left[\frac{1}{(1+(\tilde{q}_z^2+\tilde{\kappa}^2))} - \frac{u \mathcal{F}_1^2} {(2+u)}\right],
\end{align}
where $u=j+g+l+h$, and we omit the argument of hypergeometric function $\mathcal{F}_1^2=\mathcal{F}_1^2[1,\frac{2+u}{2},\frac{4+u}{2},-\frac{1}{\tilde{q}_z^2+\tilde{\kappa}^2}]$. Taking Eq.(A12), (A15) and (A17) into Eq.(A4), we recover Eq.(17) in the main text.

\section{Abrikosov's method}

In this section, we derive the explicit expressions of the relevant quantities by following Abrikosov's method \cite{Abrikosov98}. Our starting point is the $Q$-matrix given by Eq.(18) in the main text. Based on it, we can obtain the expression of conductivities from the Green's function under the eigenstate representation. Under this representation, the Hall conductivity $\sigma_{xy}$ is given by:
\begin{widetext}
\begin{align}
\sigma_{xy} &= \frac{-e^2v^2 \beta_{0}}{ 4\pi} \int \frac{dk_{z}}{2\pi} \sum_{\gamma,\gamma'}\sum_{n=1} \left[\tanh \frac{E_{n-1}^{\gamma'}\left(k_{z}\right)-\mu}{2T} - \tanh \frac{E_{n}^{\gamma}\left(k_{z}\right)-\mu}{2T} \right] \frac{\left(\chi_{n}^{-\gamma}\left(k_{z}\right) \right)^{2} \left(\chi_{n-1}^{\gamma'}\left(k_{z}\right) \right)^{2}}{\left(E_{n-1}^{+}\left(k_{z}\right)-E_{n}^{+}\left(k_{z}\right)\right)^{2}},
\end{align}
while the longitudinal conductivity is given by:
\begin{equation}
\sigma_{xx} = \frac{2e^{2}v^{2}}{T} \frac{\beta_{0}}{2\pi} \int \frac{d\omega}{2\pi} \frac{dk_{z}}{2\pi} \cosh^{-2}\frac{\omega}{2T} F\left(\omega,k_{z}\right),
\end{equation}
where
\begin{align}
F\left(\omega,k_{z}\right)&=\frac{1}{4} \sum_{\gamma,\gamma'}\sum_{n} \frac{\left(\chi_{n}^{-\gamma}\left(k_{z},B\right)\right)^{2}/\tau_{n}^{\gamma}}{\left(\omega+\mu -E_{n}^{\gamma}\left(k_{z}\right)\right)^{2}+\frac{1}{(\tau_{n}^{\gamma})^2}} \frac{\left(\chi_{n-1}^{\gamma'}\left(k_{z},B\right)\right)^{2}/\tau_{n-1}^{\gamma'}}{\left(\omega+\mu -E_{n-1}^{\gamma'}\left(k_{z}\right)\right)^{2}+\frac{1}{(\tau_{n-1}^{\gamma'})^2}}.
\end{align}
In the approach of Abrikosov, the only non-vanishing overlapping integral is from the neighbored Landau levels. Then the relaxation time is given by:
\begin{align}
\frac{1}{\tau_{n}^{\lambda}(k_{z})} = \frac{e^4}{\varepsilon_{\infty}^2}\frac{N_{i}}{v} \sum_{\zeta=\pm1} \sum_{\gamma=\pm}\int d(vk'_{z}) &\bigg\{ \frac{T}{\left(\mu-E_{n+\zeta}^{\gamma}\left(k'_{z}\right)\right)^{2}+T^{2}} \left[ \frac{(\mathcal{B}_{n,\lambda}^{n+\zeta,\gamma}\left(k_{z},k'_{z}\right))^{2}}{\beta_{0}^{2}}+(\mathcal{A}_{n,\lambda} ^{n+\zeta,\gamma}\left(k_{z},k'_{z}\right))^{2}\right] \nonumber \\
&\times \left(\ln\frac{\beta_{0}+K^{2}(k_{z},k'_{z})}{K^{2}(k_{z},k'_{z})} + \frac{K^{2}(k_{z},k'_{z})}{\beta_{0}+K^{2}(k_{z},k'_{z})}-1\right)\bigg\},
\end{align}
where $T=k_{B}T$ presents the thermal energy, $N_{i}$ is the density of charge impurities, $\mu$ is the chemical potential, and we have defined $K^{2}(k_{z},k'_{z})=\left(k_{z}-k'_{z}\right)^{2}+\kappa^{2}$ for the convenience.
\end{widetext}

In the above derivation, we have used the thermal energy $T$ to regularize the singularity of the Matsubara Green's function and account the effect of temperature around the Fermi surface. The functions $\mathcal{A}$ and $\mathcal{B}$ are defined as following:
\begin{equation}
\mathcal{A}_{n,\lambda}^{m,\gamma}\left(k_{z},k_{z}'\right)=\int dx \left(\psi_{n}^{\lambda}\right)^{\dag} x \psi_{m}^{\gamma},
\end{equation}
\begin{equation}
\mathcal{B}_{n,\lambda}^{m,\gamma}\left(k_{z},k_{z}'\right)=\int dx \left(\psi_{n}^{\lambda}\right)^{\dag} \partial_{x} \psi_{m}^{\gamma},
\end{equation}
where $\psi_{n}^{\lambda}$ is the wave function of the Landau level $E_{n}^{\lambda}$ given in Eq.(3).

\end{document}